\documentclass[epj]{svjour}
\usepackage{graphics,amsmath,color,psfrag}
\newcommand{\oper}[2]{{#1^{\phantom{\dag}}_{#2  \vphantom{{\bf i}}}}}
\newcommand{\operdag}[2]{{#1^{\dag}_{#2 \vphantom{{\bf i}}}}}
\newcommand{\opercross}[2]{{#1^{*}_{#2  \vphantom{{\bf i}}}}}

\newcommand{\integ}{{\int^\infty_{-\infty} d\varepsilon \, \nu_d(\varepsilon)}}

\begin{document}
\title{Ground State Properties of an Asymmetric Hubbard Model for Unbalanced Ultracold Fermionic Quantum Gases}
\author{T.\ Gottwald \inst{1}
\and P.\ G.\ J.\ van Dongen \inst{1}
}                     
%
%
\institute{\inst{1} KOMET 337, Institut f\"ur Physik, Johannes Gutenberg-Universit\"at, Staudingerweg 7, 55099 Mainz, Germany \\ \email{tobias.gottwald@uni-mainz.de}}
\date{Received: date / Revised version: date}
%
\abstract{In order to describe unbalanced ultracold fermionic quantum gases on optical lattices in a harmonic trap, we investigate an attractive ($U<0$) asymmetric ($t_\uparrow\neq t_\downarrow$) Hubbard model with a Zeeman-like magnetic field. In view of the model's spatial inhomogeneity, we focus in this paper on the solution at Hartree-Fock level. The Hartree-Fock Hamiltonian is diagonalized with particular emphasis on superfluid phases. For the special case of spin-independent hopping we analytically determine the number of solutions of the resulting self-consistency equations and the nature of the possible ground states at weak coupling. We present the phase diagram of the homogeneous system and numerical results for unbalanced Fermi-mixtures obtained within the local density approximation. In particular, we find a fascinating shell structure, involving normal and superfluid phases. For the general case of spin-dependent hopping we calculate the density of states and the possible superfluid phases in the ground state. In particular, we find a new magnetized superfluid phase.
\PACS{
      {03.75.Hh}{Static properties of condensates; thermodynamical, statistical, and structural properties} \and
      {39.25.+k}{Atom manipulation (scanning probe microscopy, laser cooling, etc.) (see also 82.37.Gk STM and AFM manipulations of a single molecule in physical chemistry and chemical physics; for atom manipulation in nanofabrication and processing, see 81.16.Ta)} \and
      {71.10.Fd}{Lattice fermion models (Hubbard model, etc.)}
     } 
} 
\titlerunning{Ground State Properties of an Asymmetric Hubbard Model}
\maketitle
\section{Introduction}
\label{intro}
\paragraph{Experimental background}
Ex\-peri\-ments on un\-bal\-an\-ced ul\-tra\-cold fermionic quantum gases have become of significant interest during the last year \cite{patridge:pairing,zwierlein:fermionic}. While these experiments were performed on quantum gases in the absence of an optical lattice, it will only be a matter of time before a lattice structure is superimposed on the trap potential. Since the experiments {\em without} a lattice have already been theoretically analyzed \cite{stoof:deformation,stoof:sarma}, in this paper we do the next step and investigate systems {\em with} an additional optical lattice. Such lattice systems can be described with the help of Hubbard-type models \cite{zoller:cold,greiner:mott,bloch:interfrence,oosten:quantum,bloch:nature,bloch:formation,hofstetter:hightemp,maelo:superfluidity}. Hence, since we are interested in superfluid phases with a possible imbalance in the pseudospins' occupation numbers we analyze an appropriate Hubbard model with an at\-tractive in\-ter\-action ($U<0$) and a Zeeman-like magnetic field. The symmetric and magnetic field-free version of this Hamiltonian has been intensively discussed in the seminal paper by P. Noizi\`eres and S. Schmitt-Rink \cite{noizeres:bf94sd12JE}. The pseudospin-states in this Hamiltonian represent different hyperfine states of one and the same atomic species or, alternatively, fermionic atoms with different masses. Accordingly, we also allow for a possible asymmetry in the model, i.e., for a possibly spin-dependent hopping ($t_\uparrow\neq t_\downarrow$). To deal with the spatial inhomogeneity, caused by the presence of the harmonic trap, we focus in this paper on solutions at Hartree-Fock level; possible extensions and improvements are discussed in section \ref{bd93scfguUa}. The Hartree-Fock approximation limits predictions to the region of not-too-strong (attractive) coupling between the fermions. In this paper, therefore, we restrict consideration to the moderate weak coupling regime. Clearly this regime can be realized experimentally, since interactions are tunable in ultracold quantum gases with the help of Feshbach resonances (see \cite{greiner:mott,oosten:quantum,pethick:bose}).
\paragraph{Model Hamiltonian}
A class of models, which is well suited for describing ultracold quantum gases on optical lattices, are models with a local Hubbard interaction. Here we describe only the basic ingredients of the Hamiltionian. Details of the appropriate model and its properties are discussed in section \ref{sec:2}.

In ultracold quantum gases on optical lattices, the Hubbard interaction is generated as follows.
The periodic potential induced by the coherent laser beam has the form
\begin{equation}\label{bGe64fd230}
V(\textbf{x}) = -\frac{V_0}{d} \sum^d_{i=1} \cos^2 \left( \frac{\pi x_i}{a} \right) \quad,
\end{equation}
where $V_0$ is proportional to the laser beam's intensity, $a=\frac{\lambda}{2}$ is the lattice constant and $\lambda$ is the laser beam's wavelength. In the tight binding limit, $V_0 \gg \frac{\hbar^2}{2ma^2}$, only the local overlap matrix element $U$ of the interaction contributes to the Hamiltonian. At sufficiently low temperatures ($k_{\rm B} T \ll E_{\rm gap}$, where $E_{\rm gap}$ is the band gap between the lowest two Bloch bands) an effective one-band Hubbard model may then be derived from the potential \eqref{bGe64fd230}. Apart from the on-site interaction term, which is here assumed to be attractive, the Hubbard Hamiltonian contains a nearest-neighbor hopping term. The hopping amplitudes $t_\sigma$ will in general be pseudospin-dependent, $t_\uparrow\neq t_\downarrow$, since the pseudospin index labels atoms in different hyperfine states or with different masses (see \cite{giamarchi:twocomp}). Furthermore, the particle density in the grand canonical ensemble is tuned with the help of a chemical potential, and the imbalance in the occupation numbers of the pseudospin species is regulated by a Zeeman-like magnetic field.
\paragraph{Structure of the paper}
The plan of the paper is as follows. First, in section 2, we describe the Hamiltionian, its symmetry properties and the relevant self-consistency equations for the particle density and the superfluid order parameter. Then, in sections 3 and 4, we discuss the possible solutions of the self-consistency relations for symmetric hopping, $t_\uparrow =t_\downarrow$, and the corresponding numerical results. Our most important result here is the finding of an interesting shell structure for the spatial distribution of the quantum gas in the trap. Results for the general case of asymmetric hopping, $t_\uparrow\neq t_\downarrow$, are discussed in section 5. Here we find a new magnetized superfluid phase. We end (in section 6) with a brief summary and an outlook.
\section{Properties of the Hamiltonian}
\label{sec:2}
The full grand-canonical Hamiltonian $K=H-\mu N$, describing (possibly spin-dependent) hopping of two different fermionic atomic species in a (possibly unbalanced) mixture, reads:
\begin{equation}\label{v0skm59bb1}
\begin{split}
K = &-\sum_{({\bf ij})\sigma} \oper{t}{\sigma} c^\dag_{{\bf i}\sigma} c^{\phantom{\dag}}_{{\bf j} \sigma} +U \sum_{{\bf i}} n_{{\bf i}\uparrow} n_{{\bf i}\downarrow} \\
&-\mu \sum_{{\bf i}\sigma} n_{{\bf i} \sigma} +B \sum_{{\bf i}\sigma} \sigma n_{{\bf i}\sigma} \quad.
\end{split}
\end{equation}
The various terms represent the hopping of the fermions, their local Hubbard interaction, the chemical potential and a Zeeman-like magnetic field. Alternatively, the last two terms can of course also be interpreted as a spin-dependent chemical potential $\mu_\sigma =\mu -\sigma B$. The summation label $({\bf ij})$ in the kinetic energy indicates that all nearest neighbor sites ${\bf i}$ and ${\bf j}$ are summed over, so that every bond $\langle {\bf ij}\rangle$ is counted twice. Throughout, we assume that the lattice has (hyper)cubical structure, although many symmetry properties are more generally valid for bipartite lattices. In the following, we first discuss some symmetry properties of the Hamiltonian \eqref{v0skm59bb1}, and then we derive the self-consistency equations within the Hartree-Fock approximation.
\subsection{Symmetry properties}\label{gdc6IOp0fE}
\paragraph{Chemical potential at half filling.}
The discrete symmetries of the standard Hubbard model ($B=0$ and $t_\uparrow = t_\downarrow$) are well known from the literature. In this case, the grand canonical Hamiltonian $K$ is invariant, e.g., under a spin-exchange transformation $\operdag{c}{{\bf i} \sigma} \rightarrow \operdag{c}{{\bf i}, -\sigma}$ and, at half filling on bipartite lattices, also under a general particle-hole transformation $\operdag{c}{{\bf i} \sigma} \rightarrow (-1)^{\bf i} \, \oper{c}{{\bf i} \sigma}$, where $(-1)^{\bf i}$ equals +1 for sites ${\bf i}$ on the A-sublattice and $-1$ for ${\bf i}$ on the B-sublattice. From these discrete symmetries it can be derived that, at half filling, the chemical potential is given by
\begin{equation}\label{vfg74RTde}
\mu= \frac{U}{2} \quad .
\end{equation}
In the general case (i.e., both $B \neq 0$ and $t_\uparrow \neq t_ \downarrow$) these symmetries do not exist, so that at half filling the chemical potential $\mu$ is a more complicated function of the system parameters. However, if either $B=0$ or $t_\uparrow = t_\downarrow$, equation \eqref{vfg74RTde} is still valid: For $B=0$ (but possibly $t_\uparrow \neq t_ \downarrow$), the Hamiltonian is not invariant under a spin-exchange transformation, but it is invariant under a general particle-hole transformation at half filling. Hence, \eqref{vfg74RTde} is still valid. If $t_\uparrow = t_\downarrow$ (but possibly $B\neq 0$) the Hamiltonian \eqref{v0skm59bb1} is neither invariant under a general particle-hole transformation nor under a spin-exchange transformation; nevertheless it is mapped onto itself at half filling by consecutively performing both transformations, so that \eqref{vfg74RTde} is also satisfied in this case.
\paragraph{Relation to the repulsive $\boldsymbol{U}$ model.}
In order to understand the properties of the model \eqref{v0skm59bb1} at $U<0$ and arbitrary filling, it is often helpful to map the Hamiltonian to a canonically equivalent model at $U>0$ in a magnetic field. With the use of a special particle-hole transformation, $\operdag{c}{{\bf i} \uparrow}  \rightarrow (-1)^{\bf i} \, \oper{c}{{\bf i} \uparrow}$ and $\operdag{c}{{\bf i} \downarrow} \rightarrow \operdag{c}{{\bf i} \downarrow}$, the Hamiltonian \eqref{vfg74RTde} with $U<0$ and parameters $(\mu,B)$ is mapped onto a formally identical Hamiltonian with $U>0$ and the (in general different) parameters $(\mu^\prime ,B^\prime )$. This is especially useful for $B=0$, since the new chemical potential is then given by $\mu^\prime = \frac{U}{2}$. The main advantage of the mapping is that (for symmetric hopping amplitudes) many properties of the repulsive model are well known from the literature \cite{koppe:ground}. Furthermore, with the help of a special particle-hole transformation, superfluid states in the attractive-$U$ model are mapped onto antiferromagnetic states with a staggered magnetization in the $x-y$-plane in the repulsive-$U$ model, while charge density wave states (CDW) are mapped onto antiferromagnetic states with a staggered magnetization in $z$-direction.
\subsection{Diagonalization of the Hamiltonian at Hartree-Fock level}
In order to analyze superfluid states at weak coupling, we diagonalize the Hamiltonian \eqref{v0skm59bb1} within the Hartree-Fock approximation by decoupling the interaction term
$\oper{n}{{\bf i} \uparrow} \oper{n}{{\bf i} \downarrow}$ in the usual manner:
\begin{equation}
\begin{split}
\oper{n}{{\bf i} \uparrow} \oper{n}{{\bf i} \downarrow}  \rightarrow \quad &  \oper{n}{\uparrow}  \oper{n}{{\bf i} \downarrow} +  \oper{n}{\downarrow}  \oper{n}{{\bf i} \uparrow} - \oper{n}{\uparrow} \oper{n}{\downarrow}  \\
& + \Delta \operdag{c}{{\bf i} \uparrow} \operdag{c}{{\bf i} \downarrow} + \opercross{\Delta}{} \oper{c}{{\bf i} \downarrow} \oper{c}{{\bf i} \uparrow} - | \Delta|^2 \quad ,
\end{split}
\end{equation}
where $n_\sigma \equiv \langle \oper{n}{{\bf i} \sigma} \rangle$ and $\Delta \equiv \langle \oper{c}{{\bf i} \downarrow} \oper{c}{{\bf i} \uparrow} \rangle$ are assumed to be translationally invariant thermal averages. The resulting mean-field Hamiltonian can as usual be diagonalized by a Bogoliubov-transformation. One finds
\begin{eqnarray}
\nonumber K^{\rm HF} &=& \sum_{\bf k} \left\{ E  (\oper{\varepsilon}{\textbf{k}})  - \textrm{sign} \big[ E  (\oper{\varepsilon}{\textbf{k}}) \big] \sqrt{ [E(\oper{\varepsilon}{\textbf{k}})]^2  + U^2 |\Delta|^2} \, \right\} \\
& &-\mathcal{N} U \big( |\Delta|^2 + n_\uparrow n_\downarrow \big) +\sum_{\textbf{k}\sigma} \mathcal{E} (\oper{\varepsilon}{\textbf{k}}) \bar{n}^{\phantom{\dag}}_{\textbf{k}\sigma} \quad ,
\end{eqnarray}
where $\bar{n}^{\phantom{\dag}}_{\textbf{k}\sigma}$ is the number operator of the Bogoliubov quasiparticles. Moreover, we defined $\oper{\varepsilon}{\textbf{k}}= \sum_{i=1}^d \cos k_i$ and $\mathcal{N}$ is the number of lattice sites. The spin-averaged non-superfluid particle energies $E(\varepsilon)$ and the quasiparticle energies $\mathcal{E}(\varepsilon)$ are defined as:
\begin{subequations}
\begin{align}
E(\varepsilon) =& -(t_\uparrow + t_\downarrow) \, \varepsilon - \mu + \frac{U(n_\uparrow + n_\downarrow)}{2} \\
\begin{split}
\oper{\mathcal{E}}{\sigma}(\varepsilon) =&  \; \textrm{sign} \big( E(\varepsilon) \big) \sqrt{E^2(\varepsilon) + U^2 |\Delta|^2} \; + \\
\label{vg7dRt3sl} &+ \sigma \left( (\oper{t}{\downarrow} - \oper{t}{\uparrow}) \, \varepsilon +   B +  \frac{U(n_\downarrow - n_\uparrow)}{2} \right) \quad ,
\end{split}
\end{align}
\end{subequations}
where $\sigma$ is alternatively interpreted as $\uparrow,\downarrow$ or $\pm$ in \eqref{vg7dRt3sl}. The three quantities $n_\uparrow$, $n_\downarrow$ and $\Delta$ can also be determined from their definitions as thermal averages. Suppressing all details, we just mention as a result that, at zero temperature, the following three self-consistency equations have to be satisfied in the thermodynamic limit:
\begin{subequations}
\begin{align}
\label{bu7dfT53vk}
n_\uparrow =& \frac{1}{2} \left[ 1 + \int_{-\infty}^\infty d\varepsilon \, \nu_d (\varepsilon)  \frac{\big| E(\varepsilon) \big| \, \textrm{sign} \big( -\mathcal{E}_\uparrow (\varepsilon) \big)}{\sqrt{E^2(\varepsilon) +U^2 |\Delta|^2}} \right]    \\
\label{bhYTD32sgT}
n_\downarrow =& \frac{1}{2} \left[ 1+ \int_{-\infty}^\infty d\varepsilon \, \nu_d (\varepsilon)  \frac{ \big| E(\varepsilon) \big| \, \textrm{sign} \big( -\mathcal{E}_\downarrow (\varepsilon) \big)}{\sqrt{E^2(\varepsilon) +U^2 |\Delta|^2}} \right]    \\
\begin{split}
\label{o9d3acyU8fd}
\Delta =& \frac{U\Delta}{2} \int_{-\infty}^\infty d\varepsilon \, \nu_d (\varepsilon)  \frac{\textrm{sign} \big( E(\varepsilon) \big) }{\sqrt{E^2(\varepsilon) +U^2 |\Delta|^2}} \; \times \\
& \times \left[ \Theta \big( -\mathcal{E}_\uparrow (\varepsilon) \big) + \Theta \big( -\mathcal{E}_\downarrow (\varepsilon) \big) -1 \right] \quad ,
\end{split}
\end{align}
\end{subequations}
where $\nu_d (\varepsilon)$ is the $d$-dimensional density of states of the interaction-free Hubbard model. Since particle-hole symmetry is lost if both $B \neq 0$ and $t_\uparrow \neq t_\downarrow$ (see section \ref{gdc6IOp0fE}), we consider in the following only special cases with either $B=0$ or $t_\uparrow = t_\downarrow$.
\section{Properties of the self-consistency equations for symmetric hopping ($\boldmath t_{\boldsymbol{\uparrow}} = t_{\boldsymbol{\downarrow}}\equiv t$)}\label{Tbji9dEvR2}
Since the coupled equations \eqref{bu7dfT53vk} - \eqref{o9d3acyU8fd} may have more than one solution, we show analytically that the number of solutions at fixed $\{ U,t,\mu,B \}$ is between one and three. If more than one solution is found, the one with the lowest grand potential is thermodynamically stable.
\subsection{Particle numbers at fixed $\boldsymbol{\Delta}$}\label{Uirvb6dR3s}
\paragraph{Integral equations as a mapping.}
In order to determine the number of solutions of the coupled equations \eqref{bu7dfT53vk} - \eqref{o9d3acyU8fd}, we first consider Equations \eqref{bu7dfT53vk} and \eqref{bhYTD32sgT} separately at fixed order parameter $\Delta$. We show that, at fixed $\Delta$, there is a unique solution of \eqref{bu7dfT53vk} and \eqref{bhYTD32sgT} by using Banach's fixed point theorem. As vectors we use ${\bf n}_i := (n_{i\uparrow}, n_{i\downarrow})^{\rm T}$, where $i$ denotes different points in the pa\-ra\-me\-ter space, i.e., different values of $\{ U,t,\mu,B \}$. The integrations of \eqref{bu7dfT53vk} and \eqref{bhYTD32sgT} can be interpreted as a mapping $\mathcal{I}$ from the parameter space onto itself. Furthermore, in order to apply Banach's theorem, we consider the maximum metric $||{\bf n}_i - {\bf n}_j ||_\infty = {\rm max}_\sigma \{|n_{i,\sigma}- n_{j,\sigma} | \}$. To prove uniqueness it is sufficient to show that a number $0 < q <1$ exists, so that for all ${\bf n} \in [0,1]^2$:
\begin{equation}\label{vgYtsE3v58}
|| \mathcal{I} ({\bf n +\delta n}) - \mathcal{I} {\bf n}||_\infty \leq q \, ||{\bf \delta n}||_\infty 
\end{equation}
holds, where ${\bf \delta n}$ is a small variation in parameter space. Hence, it suffices to show that the mapping $\mathcal{I}$ is a contraction. The solution of \eqref{bu7dfT53vk} and \eqref{bhYTD32sgT} is then given by the unique fixed point of the equation ${\bf n}_0 = \mathcal{I}{\bf n}_0$.
\paragraph{Convergence at small interaction strength.}
For small variations ${\delta \bf n}$ it can be shown that the mapping $\mathcal{I}$ is a contraction for all points ${\bf n}$ in parameter space. In order to prove that Equation \eqref{vgYtsE3v58} is indeed satisfied at every point ${\bf n}$, we use a Taylor-expansion of $|\mathcal{I} ( n_\sigma +\delta  n_\sigma) - \mathcal{I} n_\sigma|$ in the parameter ${\delta \bf n}$ and some estimates, whose validity depends upon the relative interaction $\frac{|U|}{t}$ being weak. In doing so, we must distinguish two cases, namely that the functions $\mathcal{E}_\sigma (\varepsilon)$ and $E(\varepsilon)$ change sign at the {\em same} value or at {\em different} values of $\varepsilon$. In the first case the Taylor-expansion yields as a criterion for the validity of \eqref{vgYtsE3v58} with $0<q<1$:
\begin{equation}\label{hu8dTes5}
\frac{|U|}{t} < \frac{2}{\nu_d (\varepsilon_{\rm max})}
\end{equation}
and in the second case it yields:
\begin{equation}\label{Hywqa7vOP}
\frac{|U|}{t} < \frac{1}{2 \, \nu_d (\varepsilon_{\rm max})} \quad .
\end{equation}
Here $\varepsilon_{\rm max}$ is defined as the energy, for which the non-interacting density of states is maximal. Since the condition \eqref{Hywqa7vOP} is more restrictive than \eqref{hu8dTes5}, Equation \eqref{Hywqa7vOP} guarantees that the mapping $\mathcal{I}$ is a contraction in the whole parameter space at sufficiently weak coupling. Note that this proof does not work if the density of states diverges at a van Hove singularity. For a simple cubic lattice, Equation \eqref{Hywqa7vOP} yields $\frac{|U|}{t} < 1.65$ for $\mathcal{I}$ to be a contraction.
\subsection{Variation of the order parameter $\boldsymbol{\Delta}$}
In order to solve the coupled Equations \eqref{bu7dfT53vk} - \eqref{o9d3acyU8fd} we also have to discuss the properties of \eqref{o9d3acyU8fd} explicitly. Equation \eqref{o9d3acyU8fd} is trivially solved by $\Delta=0$, which corresponds to a non-superfluid phase. For $\Delta \neq 0$ and symmetric hopping amplitudes it is useful to distinguish between $|B| <|U \Delta|$, where $n_\uparrow = n_\downarrow$ and Equation \eqref{o9d3acyU8fd} reduces to
\begin{equation}
\label{AbhYU456d7}
0 = \frac{|U|}{2} \integ \frac{1}{ \sqrt {E^2 (\varepsilon) + U^2 |\Delta|^2}} -1 \quad ,
\end{equation}
and $|B| >|U \Delta|$, where \eqref{o9d3acyU8fd} may be rewritten as:
\begin{equation}
\label{DGtbnsw6vft}
0 = \frac{|U|}{2} \integ \frac{ \Theta \big( \mathcal{E}_\uparrow (\varepsilon) \cdot \mathcal{E}_\downarrow (\varepsilon) \big) }{ \sqrt {E^2 (\varepsilon) + U^2 |\Delta|^2}} -1 \quad .
\end{equation}
Since, from section \ref{Uirvb6dR3s}, we know that there is only one solution for $n_\uparrow$ and $n_\downarrow$, we are able to do a one-dimensional parameter scan through all possible $|\Delta|$-values in order to find the solutions of \eqref{o9d3acyU8fd}, using Equations \eqref{DGtbnsw6vft} for $|\Delta| < \left| \frac{B}{U} \right|$ and \eqref{AbhYU456d7} for $|\Delta| > \left| \frac{B}{U} \right|$.
Since the integral on the right in \eqref{AbhYU456d7} decreases with increasing $|\Delta|$ if $|\Delta| > \left| \frac{B}{U} \right|$, there is at most one solution in this interval. Similarly, there is also at most one solution for $0< |\Delta| < \left| \frac{B}{U} \right|$, since the integral on the right in \eqref{DGtbnsw6vft} in general increases with $|\Delta|$ in this case, so that the residual is monotonic on both sides of $\left| \frac{B}{U} \right|$. As a consequence, there exist at most three solutions for $|\Delta|$, namely the trivial solution $|\Delta| = 0$ and possibly two non-trivial solutions on either side of $\left| \frac{B}{U} \right|$. Note that the modulus of $\Delta = \langle \oper{c}{{\bf i} \downarrow} \oper{c}{{\bf i} \uparrow} \rangle$ is rigorously bounded from above by $|\Delta| < \frac{1}{2}$. 

\subsection{Possible phases occurring at Hartree-Fock level}\label{vf5D2sO0}
Since, for $|\Delta| > \left| \frac{B}{U} \right|$, the magnetization is rigorously zero ($n_\uparrow = n_\downarrow$) and there are at most three non-equivalent solutions of the coupled Equations \eqref{bu7dfT53vk} - \eqref{o9d3acyU8fd}, we find that (for $B \neq 0$) there can in principle be up to three competing phases:
\begin{itemize}
\item a non-superfluid magnetized phase ($\Delta = 0$)
\item a superfluid magnetized phase ($0 < |\Delta| < \left| \frac{B}{U} \right|$)
\item a superfluid non-magnetized phase ($\left| \frac{B}{U} \right| < |\Delta|$)\; .
\end{itemize}
For $B=0$ it is well known that the superfluid non-magnetized solution has the lowest grand potential. Since this is not necessarily true for $B \neq 0$, phase transitions will generally be of first order, also at Hartree-Fock level. We should also mention that there is no thermodynamically stable superfluid magnetized phase at $T=0$ and $t_\uparrow = t_\downarrow$, since that type of solution is always a local maximum of the grand potential which can be shown with elementary mathematical methods.
\section{Numerical results for symmetric hopping}\label{bg9gsWJI4ER}
\begin{figure*}[!h]
\begin{center}
  \begin{minipage}{58mm}
   \psfrag{B}[r][r]{\huge $B \;$}
   \psfrag{Delta}[l][l]{\huge $\Delta$}
   \psfrag{mu}[l][l]{\huge $\mu$}
   \resizebox{0.9\textwidth}{!}{
   \includegraphics{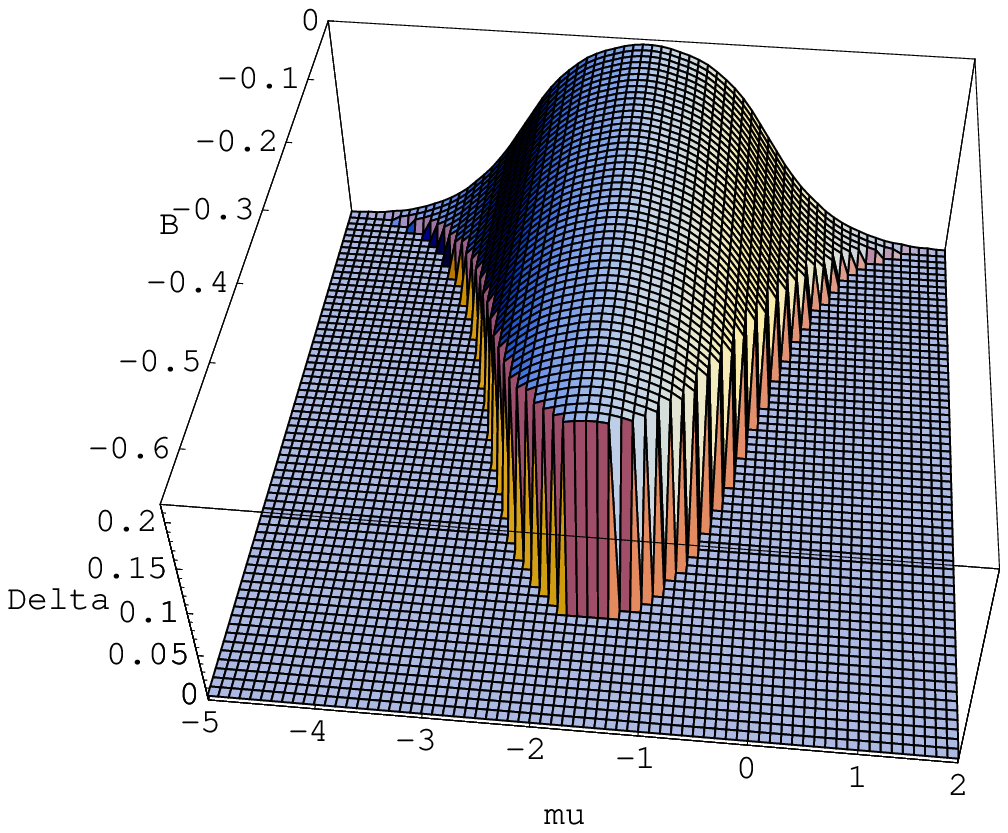}}
  \end{minipage}
  \begin{minipage}{58mm}
   \psfrag{m}[r][r]{\huge $M$}
   \psfrag{B}[r][r]{\huge $B \;$}
   \psfrag{mu}[l][l]{\huge $\mu$}
   \resizebox{0.9\textwidth}{!}{
   \includegraphics{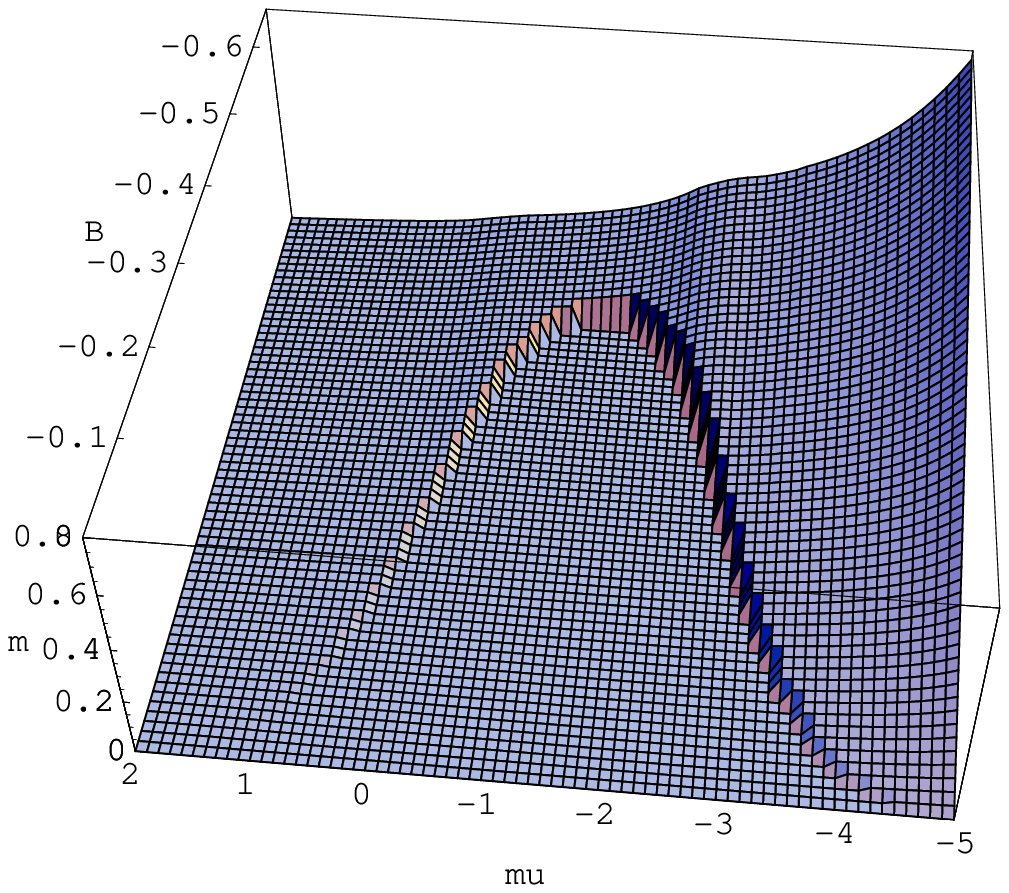}}
  \end{minipage}
  \begin{minipage}{58mm}
   \psfrag{n}[r][r]{\huge $n \;$}
   \psfrag{mu}[r][r]{\huge $\mu$}
   \psfrag{B}[r][r]{\huge $B \quad$}
   \resizebox{0.9\textwidth}{!}{
   \includegraphics{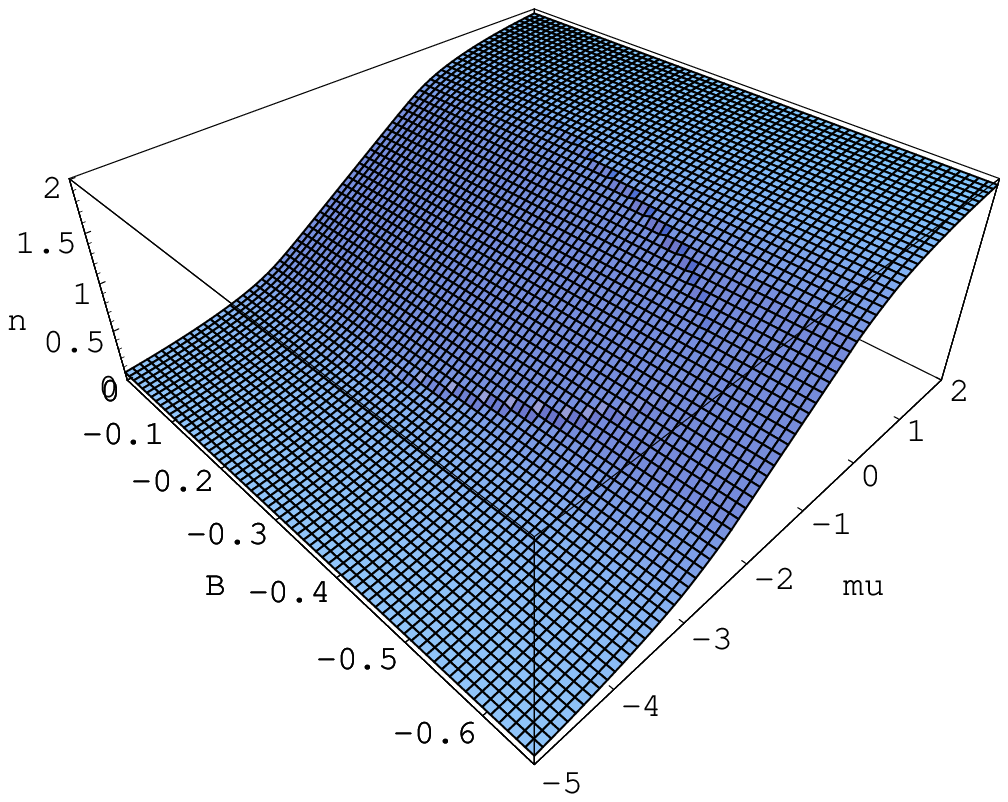}}
  \end{minipage}
\caption{Phase diagram of the homogeneous system at $T=0$, $t=1$ and $U=-3$. The parameters $n$, $M$ and $\Delta$ are shown as a function on $\mu$ and $B$. The jump in the superfluid order parameter $\Delta$ and the magnetization $M$ indicates the first order phase transition. The relative jump of the particle number $n$ is smaller and is therefore not adequate as a signature of superfluidity.}\label{T5bbf09dw}
\end{center}
\end{figure*}

\paragraph{Numerical method.}
With the knowledge of the properties of the self-consistency equations \eqref{bu7dfT53vk} - \eqref{o9d3acyU8fd}, established in section \ref{Tbji9dEvR2}, it is possible to do accurate numerical calculations for the phases occurring in the ground state. Specifically, we first determine all possible solutions of Equations \eqref{bu7dfT53vk} - \eqref{o9d3acyU8fd}, and then identify the thermodynamically stable solution by comparing the respective numerical values of the grand potentials. The various solutions of the selfconsistency equations are determined by performing a scan over $|\Delta|$-values between $0$ and $\frac{1}{2}$ in order to find the roots of \eqref{o9d3acyU8fd}. The parameters $n_\uparrow$ and $n_\downarrow$ are varied self-consistently at each step by successive approximation; this method is guaranteed to work on account of Banach's fixed point theorem (see section \ref{Uirvb6dR3s}). The phase diagram of the homogeneous system is presented in Figure \ref{T5bbf09dw}. The parameters $n$, $M$ and $\Delta$ are presented as a function of $\mu$ and $B$ at fixed $U$ and $t$. At $B=0$ the system is always superfluid and the parameters $n$, $M$ and $\Delta$ do not depend on $B$ at fixed $\mu$ until a critical magnetic field $B_C$, where superfluidity breks down and magnetization occurs. Note that this first order phase transition generally takes place at $|B_C| < |U \Delta_{B=0}|$. The superfluid order parameter is of course maximal at half filling ($\mu= \frac{U}{2}$).
\begin{figure*}[!h]
 \begin{center}
   \begin{minipage}{58mm}
   \psfrag{c}[l][l]{\huge $n \quad$}
   \psfrag{a}[r][r]{\huge x}
   \psfrag{b}[l][l]{\huge z}
   \resizebox{0.9\textwidth}{!}{
   \includegraphics{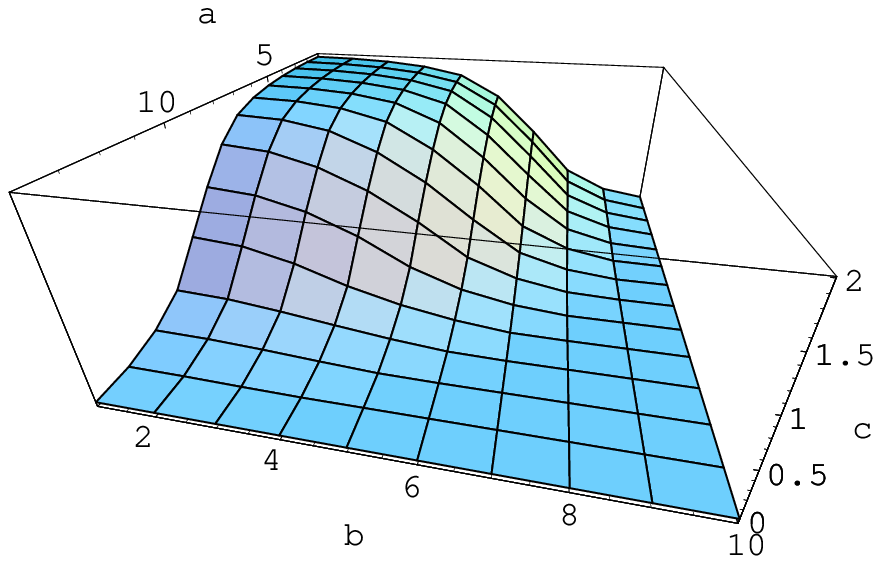}}
  \end{minipage}
  \begin{minipage}{58mm}
   \psfrag{c}[r][r]{\huge $M$ $\quad$}
   \psfrag{a}[r][r]{\huge x}
   \psfrag{b}[l][l]{\huge z}
   \resizebox{0.9\textwidth}{!}{
   \includegraphics{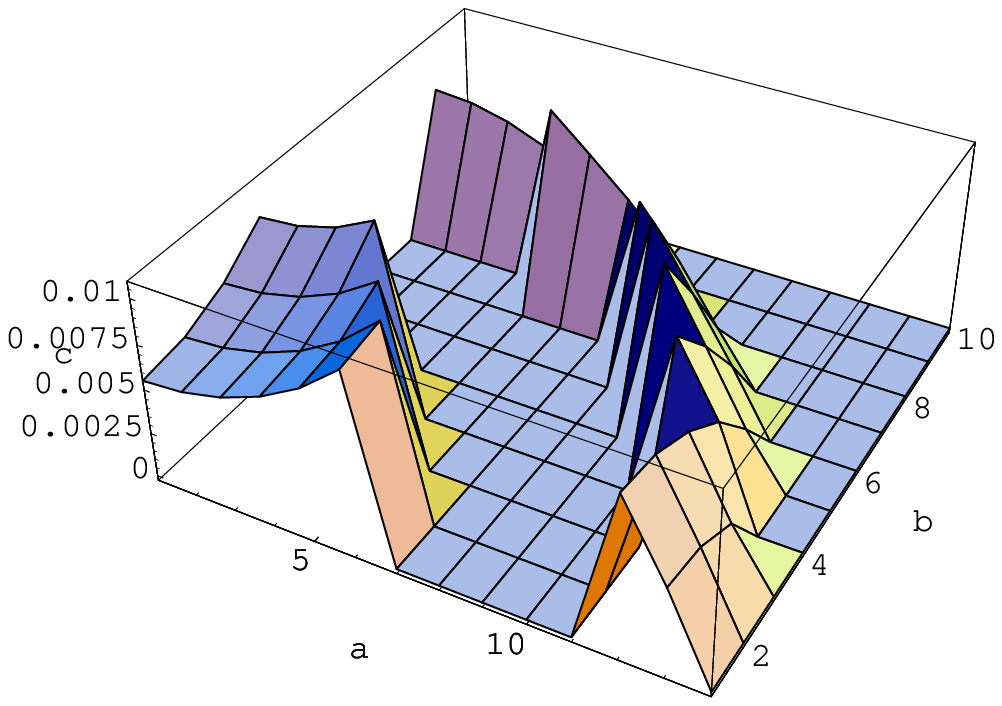}}
  \end{minipage}
  \begin{minipage}{58mm}
   \psfrag{c}[c][c]{\huge $|\Delta| \qquad  $}
   \psfrag{a}[r][r]{\huge x}
   \psfrag{b}[l][l]{\huge z}
   \resizebox{0.9\textwidth}{!}{
   \includegraphics{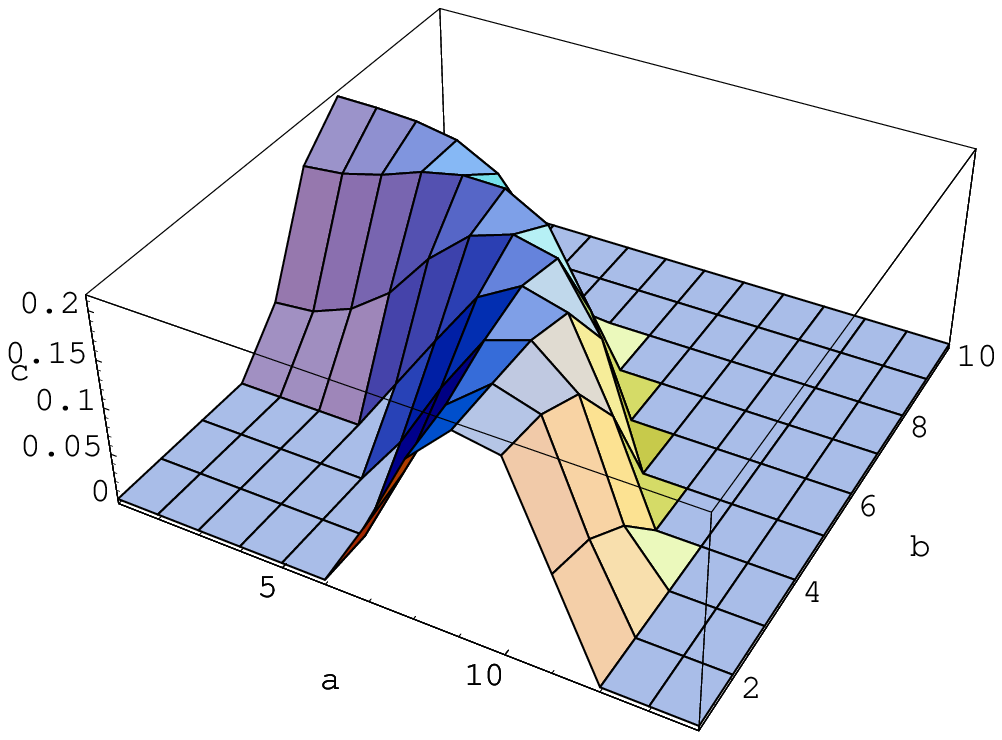}}
  \end{minipage}
 \end{center}
 \caption{Spatial distribution of the parameters $n$, $M=n_\uparrow -n_\downarrow$ and $|\Delta|$ in the x-z-plane of a simple cubic lattice at trap parameters: $\{ U=-3, \, t=1, \, \mu_0=2, \, B=-0.1, \, \Omega=\textrm{Diag}(0.05,0.3,0.1) \}$.}\label{bH7f4Rtco}
\end{figure*}
\begin{figure*}[!h]
 \begin{center}
  \begin{minipage}{58mm}
   \psfrag{c}[l][l]{\huge $n \quad$}
   \psfrag{a}[r][r]{\huge x}
   \psfrag{b}[l][l]{\huge z}
   \resizebox{0.9\textwidth}{!}{
   \includegraphics{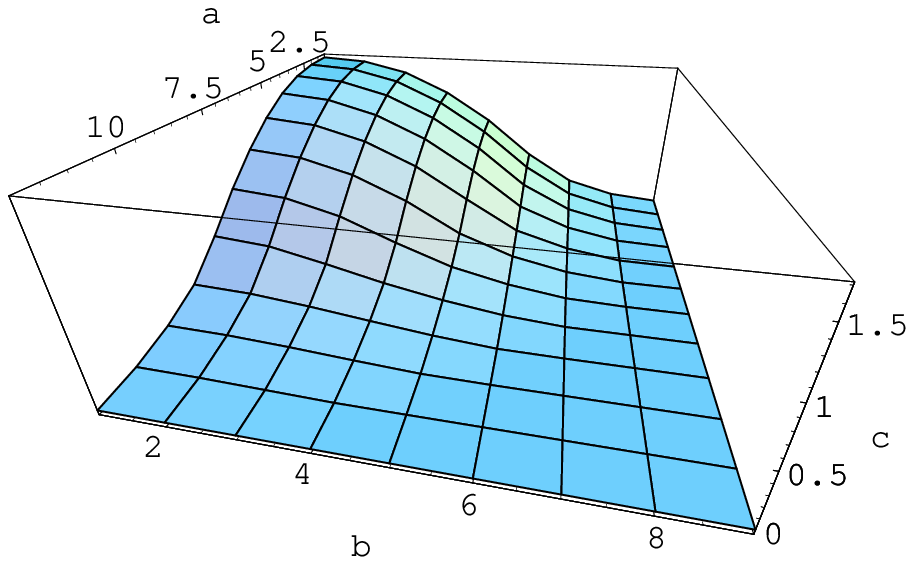}}
  \end{minipage}
  \begin{minipage}{58mm}
   \psfrag{c}[r][r]{\huge $M$ $\quad$}
   \psfrag{a}[r][r]{\huge x}
   \psfrag{b}[l][l]{\huge z}
   \resizebox{0.9\textwidth}{!}{
   \includegraphics{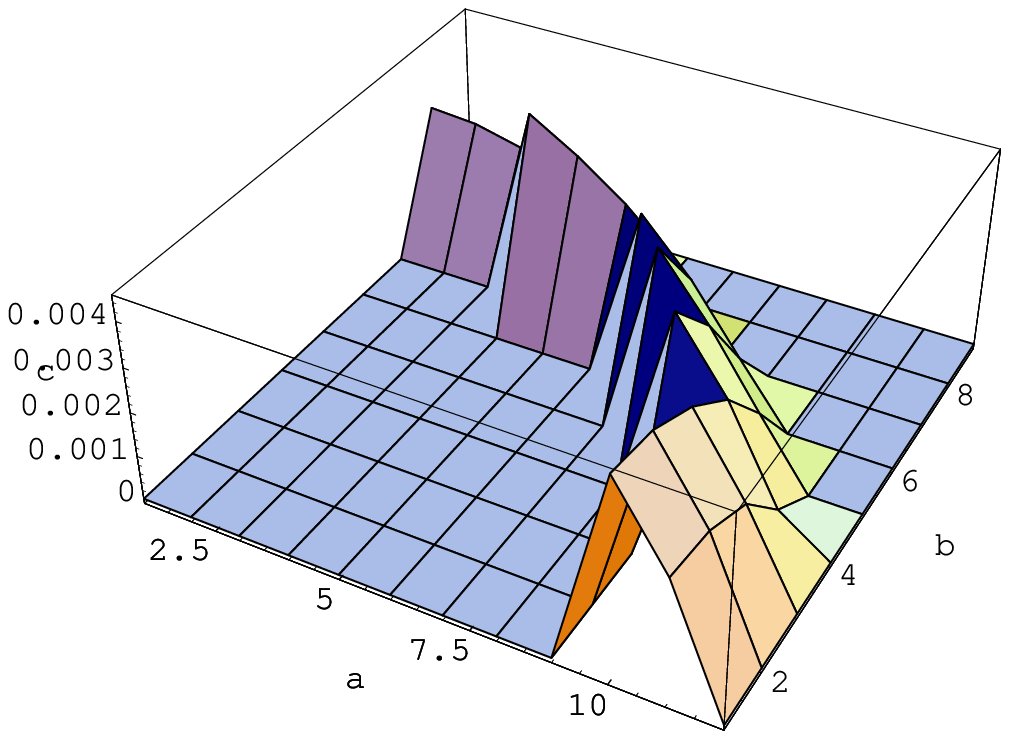}}
  \end{minipage}
  \begin{minipage}{58mm}
   \psfrag{c}[c][c]{\huge $|\Delta| \qquad  $}
   \psfrag{a}[r][r]{\huge x}
   \psfrag{b}[l][l]{\huge z}
   \resizebox{0.9\textwidth}{!}{
   \includegraphics{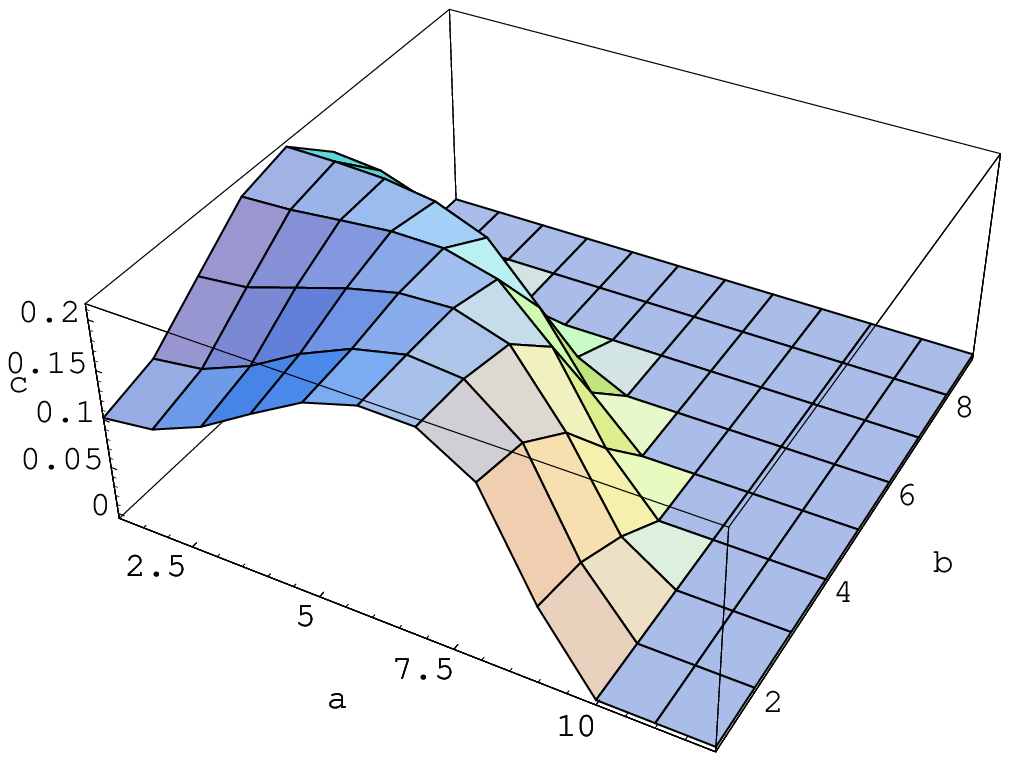}}
  \end{minipage}
 \end{center}
 \caption{Spatial distribution of the parameters $n$, $M=n_\uparrow -n_\downarrow$ and $|\Delta|$ in the x-z-plane of a simple cubic lattice at trap parameters: $\{ U=-3, \, t=1, \, \mu_0=0, \, B=-0.05, \, \Omega=\textrm{Diag}(0.05,0.3,0.1) \}$.}\label{Lpft6eRt3s}
\end{figure*}
\begin{figure*}[!h]
 \begin{center}
  \begin{minipage}{58mm}
   \psfrag{c}[l][l]{\huge $n \quad$}
   \psfrag{a}[r][r]{\huge x}
   \psfrag{b}[l][l]{\huge z}
   \resizebox{0.9\textwidth}{!}{
   \includegraphics{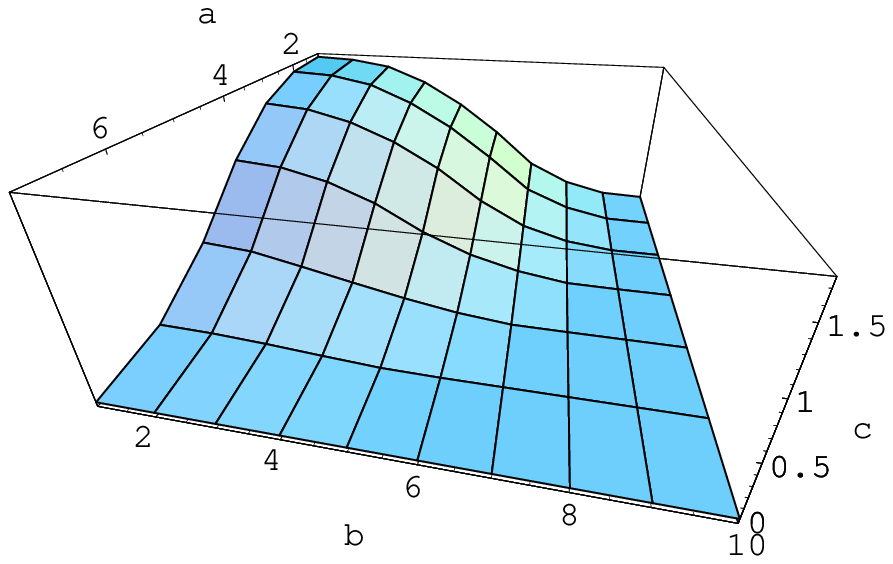}}
  \end{minipage}
 \hspace{1.5cm}
  \begin{minipage}{58mm}
   \psfrag{c}[r][r]{\huge $M$ $\quad$}
   \psfrag{a}[r][r]{\huge x}
   \psfrag{b}[l][l]{\huge z}
   \resizebox{0.9\textwidth}{!}{
   \includegraphics{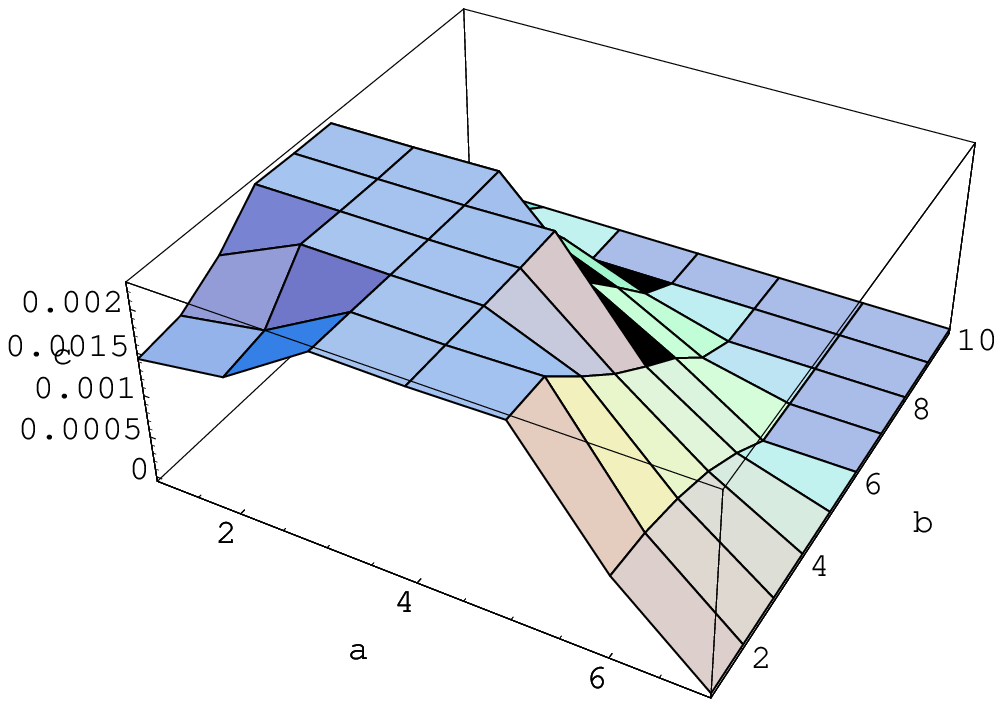}}
  \end{minipage}
 \end{center}
 \caption{Spatial distribution of the parameters $n$ and $M=n_\uparrow -n_\downarrow$ in the x-z-plane of a simple cubic lattice at trap parameters: $\{ U=-2, \, t=1, \, \mu_0=1, \, B=-0.01, \, \Omega=\textrm{Diag}(0.2,0.3,0.1) \}$. The order parameter $|\Delta|$ vanishes in the whole system and is not shown.}\label{NthY74dsU}
\end{figure*}
\paragraph{Numerical results within the LDA.}
In order to formulate predictions for experiments on ultracold quantum gases we also have to treat the parabolic potential arising from the magneto-optical trap. In Hubbard model language this potential may be described as:
\begin{equation}
H_{\rm Trap} = \sum_{\bf i \sigma} ({\bf i} \cdot \Omega \cdot {\bf i}) \oper{n}{\bf i \sigma} \quad ,
\end{equation}
where $\Omega$ is a real, symmetric, positive definite matrix. We assume the trapping potential to be spin-independent. The trapping potential is treated within the local density approximation (LDA), meaning that local quantities at site ${\bf i}$ are determined from the Hartree-Fock selfconsistency equations corresponding to the effective chemical potential $\mu({\bf i}) = \mu_0 -({\bf i} \cdot \Omega \cdot {\bf i})$; the global chemical potential $\mu_0$ is then tuned so as to reproduce the desired total number of particles in the trap. 

The numerical results, obtained from the Hartree-Fock-LDA approximation, are presented in  Figures \ref{bH7f4Rtco} - \ref{NthY74dsU}, which show the spatial distribution of the occupation number $n \equiv n_\uparrow + n_\downarrow$, the spatial distribution of polarization $M \equiv n_\uparrow - n_\downarrow$ and the spatial distribution of the superfluid order parameter's absolute value $|\Delta|$, respectively. In accordance with the ellipsoidal geometry of the trap, we find a fascinating and intuitively extremely plausible shell structure for the spatial distribution of the various densities and the local order parameter. In Figure \ref{bH7f4Rtco} there is a normal phase in the center of the trap surrounded by a superfluid non-magnetized shell and a normal phase in the outer region of the trap. Apparently, the normal phase in the center of the trap is stabilized by the Zeeman-like magnetic field and reacts very sensitively to changes in this field. This can be seen from Figure \ref{Lpft6eRt3s}, where the (absolute value of the) magnetic field is slightly lower than in Figure \ref{bH7f4Rtco}; also the global particle number (chemical potential $\mu_0$) is slightly different. These changes suffice to cause the disappearance of the non-superfluid core. By enhancing the stiffness of the trap potential in the $x$- and $z$-directions or slightly decreasing the interaction strength $|U|$, it is possible to suppress superfluidity alltogether: There is no superfluid phase in Figure \ref{NthY74dsU}.
\paragraph{Experimental consequences.}
In experiments, the superfluid order parameter $\Delta$ cannot be observed directly. On the basis of the results, presented above, it seems that little information of use concerning superfluidity can be deduced from the total particle number $n_\uparrow + n_\downarrow$. However, the other observable considered above, i.e., the local polarization $M = n_\uparrow - n_\downarrow$, might in fact be quite helpful in identifying the shell-structure in the trap as an experimental signature for superfluidity in unbalanced Fermi-mixtures. The observable $M=n_\uparrow - n_\downarrow$ can be resolved spatially with the use of in-situ imaging \cite{patridge:pairing,zwierlein:fermionic}.
\paragraph{Comparison with optical lattice-free systems.}
Systems {\em without optical lattices} have been intensely discussed in the literature, both from a theoretical \cite{stoof:deformation,stoof:sarma} as well as from an experimental point of view \cite{patridge:pairing,zwierlein:fermionic}. Important {\em similarities} between systems with and without optical lattices are, first, the occurrence of superfluidity below a critical temperature $T_C$ and secondly, the observation of a shell structure containing normal and superfluid rings. Moreover, there are also several {\em differences} between these classes of systems, which will now be highlightned.

In optical lattice-free systems it is found, in contrast to our results, that, if superfluidity occurs, the atoms in the center of the trap are always superfluid. This difference arises from the fact that, in the one-band situation discussed in this paper, the quasimomentum is bounded from above by $k_i \leq 2 \pi$, whereas in optical lattice-free systems there is no upper bound for the components of the momentum vector. We have shown that superfluid magnetized phases occurring in optical lattice-free systems as demonstrated, e.g., by Parish et al. \cite{parish:finite:6Ge,parish:polarized:v8E} do not occur in systems, considered in this work (see section \ref{vf5D2sO0}).

Deformations of the shell structure in the optical lattice-free system, discussed by H. Stoof in \cite{stoof:deformation}, seem to arise from a surface tension between a superfluid core and a normal state, which is not discussed in our work, where we consider systems described by discrete lattice vectors.

Moreover, in recent publications \cite{sheehy:b84F3oa} FFLO phases in lattice-free systems are discussed. At present it seems not possible to combine this formalism, used for spatially homo\-geneous (or periodic) potentials, with the LDA-ap\-proach, used in this paper for spatially inhomogeneous systems, since the FFLO-ground state is spatially inhomogeneous by itself. For this reason, and since FFLO phases were not yet observed experimentally, we disregard FFLO phases in this paper.
\paragraph{Trapping potential-free systems.}
In recent literature \cite{iskin:bgu93dbE}, trapping potential-free systems have been discussed in the weak as well as in the strong coupling regime. In the weak coupling regime we obtain the phase diagram in the grand canonical ensemble, presented in Figure \ref{T5bbf09dw}. Since we find a first order phase transition between a superfluid non-magnetized phase and a normal phase, we can formulate predictions for systems with fixed particle numbers $n_\uparrow$ and $n_\downarrow$ instead of fixed conjugate variables $\mu$ and $B$: For non-magnetized systems $n_\uparrow = n_\downarrow$ the system is completely superfluid. For a population imbalance, $n_\uparrow \neq n_\downarrow$, the following case differentiation has to be done:
\begin{enumerate}
\item If in the normal phase (i.e. assuming $\Delta=0$) the particle numbers $n_\uparrow$ and $n_\downarrow$ can be realized (within the grand canonical ensemble) by choosing definite values $\mu$ and $B$ for the chemical potential and the magnetic field, respectively, then this normal phase is also the ground state.
\item If in the normal phase there exists no such combination of ($\mu , B$)-values, then phase separation occurs in the ground state. The superfluid and the normal fraction have to be determined by a Maxwell construction.
\end{enumerate}
As mentioned in section \ref{bg9gsWJI4ER} a superfluid magnetized phase does not exist, at least not in the weak coupling limit.
\section{Results for asymmetric hopping}
In this section we discuss the properties of the Hartree-Fock Hamiltonian for superfluid states with asymmetric hopping ($t_\uparrow\neq t_\downarrow$). For clarity and convenience, we focus on translationally invariant solutions ($\Omega =0$) for $B=0$.
\subsection{Charge density wave states and the repulsive $\boldsymbol{U}$ model}
As emphasized before, we are interested in (and, hence, focussing on) superfluid phases in the negative-$U$ Hubbard model, disregarding possible competing phases, which seem to be irrelevant for experiments on ultracold quantum gases (see for example \cite{hofstetter:hightemp,maelo:superfluidity,orso:sound,yamada:novel,shly:superfluid,reddy:asymmetric}). The issue of possible competing phases becomes particularly relevant for $B=0$ and {\em half-filling}, since in this case charge density wave (CDW) states are known to be degenerate with s-wave superfluidity if $t_\uparrow = t_\downarrow$. This is immediately clear from a special particle-hole transformation from the negative-$U$ to the positive-$U$ Hubbard model, both at half-filling and $B=0$, since under this transformation s-wave and CDW states are mapped onto the various directions of the staggered magnetization, which are energetically degenerate due to the Hubbard model's $SU(2)$-symmetry in the spin sector. Away from half-filling, however, s-wave superfluidity is stabilized with respect to the CDW phase. 

For moderately asymmetric hopping ($t_\uparrow \simeq t_\downarrow$) the situation is similar: Exactly at half-filling the CDW phase is energetically even somewhat lower than s-wave superfluidity, so that one expects the CDW state to dominate in the phase diagram in a narrow but finite strip around half-filling. We conclude that the negative-$U$ Hubbard model, considered here, predicts phases near half-filling, which at present seem to be of little experimental interest. For this reason we concentrate in the following on band fillings, which deviate significantly from half-filling, so that the superfluid phases, considered here, can safely be assumed to be stable \cite{koppe:ground}.
\subsection{Hartree-Fock density of states}
The absence of symmetry in the hopping amplitudes ($t_\uparrow \neq t_\downarrow$) clearly also influences the (Hartree-Fock) density of states, which is defined as:
\begin{equation}
\label{DOStg1}
\nu^{\rm HF}_{d,\sigma} (\mathcal{E}) = \integ \delta \big( \mathcal{E} - \mathcal{E}_\sigma (\varepsilon) \big) \; .
\end{equation}
The asymmetric hopping now causes ad\-ditio\-nal sin\-gu\-lari\-ties (jumps) in the spin-dependent densities of states. If we assume $0 < t_\downarrow < t_\uparrow$, the following changes occur:
\begin{itemize}
\item The inverse square-root divergence of the density of states at the superfluid gap in $\nu_{d,\uparrow}$ is replaced by a discontinuity (jump).
\item The DOS $\nu_{d,\downarrow}$ retains its inverse square-root divergence but now displays two additional discontinuities within the band (one below and one above the gap).
\item The superfluid gap in $\nu_{d,\downarrow}$ for $t_\uparrow \neq t_\downarrow$ is smaller than the corresponding value $2|U\Delta|$ for $t_\uparrow = t_\downarrow$, while the gap remains unchanged in $\nu_{d,\uparrow}$.
\end{itemize}
The normalization of $\nu_{d,\sigma}$ is obviously not affected. 

\begin{figure}[!h]
 \begin{center}
 \psfrag{a}[l][l]{$\mathcal{E}$}
 \psfrag{b}[c][c]{$\nu_{3,\uparrow}(\mathcal{E})$}
 \resizebox{0.9\columnwidth}{!}{
  \includegraphics{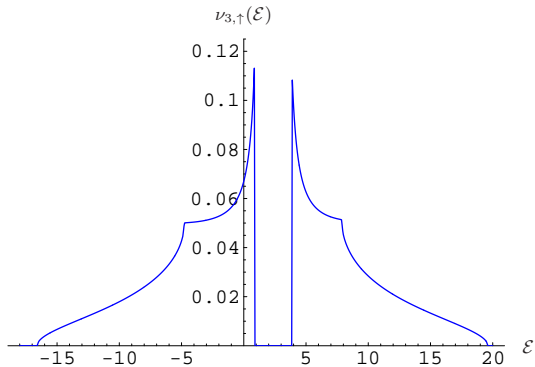}}
 \end{center}
\caption{Hartree-Fock DOS of the spin species with the greater hopping amplitude. Instead of a square-root singularity there is a jump at the border of the superfluid gap.}\label{Pvfd7Easd0}
\end{figure}

\begin{figure}[!h]
 \begin{center}
 \psfrag{a}[l][l]{$\mathcal{E}$}
 \psfrag{b}[c][c]{$\nu_{3,\downarrow}(\mathcal{E})$}
 \resizebox{0.9\columnwidth}{!}{
  \includegraphics{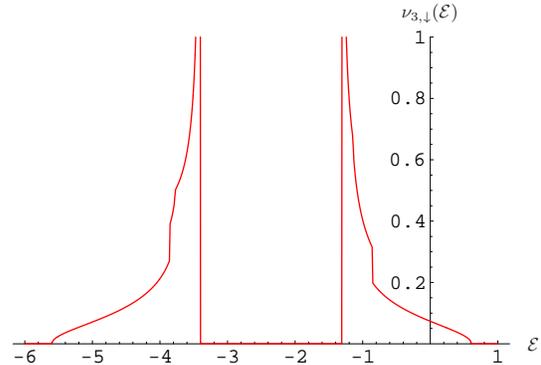}}
 \end{center}
\caption{Hartree-Fock DOS of the spin-species with the lower hopping-amplitude. Note that an additional jump appears within the band, and that the superfluid gap is smaller than $2|U\Delta|$.}\label{Afy73d8902}
\end{figure}

The densities of states for both spin species, as determined from Equations \eqref{DOStg1} and \eqref{vg7dRt3sl}, are shown in Figures \ref{Pvfd7Easd0} and \ref{Afy73d8902} for the parameter values $\{ U=-5 , \; t_\uparrow = 3, \; t_\downarrow= 0.5, \; \mu= -2, \; n_\uparrow= 0.7, \; n_\downarrow=0.3 \}$ and a three-dimensional simple cubic lattice. These parameter values are only used to illustrate the mentioned effects on the DOS; they do not necessarily represent a solution of the self-consistency equations \eqref{bu7dfT53vk} - \eqref{o9d3acyU8fd}. The spin-dependent size of the gap, as seen in Figures \ref{Pvfd7Easd0} and \ref{Afy73d8902}, can of course also be observed experimentally.
\subsection{Phases in the ground state}
In this section we present the results of the numerical solutions of \eqref{bu7dfT53vk} - \eqref{o9d3acyU8fd} for translationally invariant simple cubic lattices at fixed particle numbers $n=n_\uparrow + n_\downarrow$ and $B=0$. The proof presented in section \ref{Uirvb6dR3s} is also valid for asymmetric hopping ($t_\uparrow\neq t_\downarrow$), albeit for smaller values of the dimensionless interaction parameter $\frac{|U|}{{\rm max}_\sigma \{t_\sigma\}}$ than for systems with symmetric hopping. This is due to the fact that the quasiparticle energy function $\mathcal{E}(\varepsilon)$ in \eqref{vg7dRt3sl} is not a monotonic function in $\varepsilon$ for the case of asymmetric hopping. Numerical calculations show that,  for $B=0$ and $t_\uparrow \neq t_\downarrow$, Equation \eqref{o9d3acyU8fd} has at most one solution with $\Delta \neq 0$. If two solutions (one with $\Delta = 0$ and one with $\Delta \neq 0$) exist, we find that the superfluid phase always minimizes the free energy. 

\begin{figure*}
 \begin{center}
  \begin{minipage}{58mm}
   \psfrag{a}[l][l]{\huge $U$}
   \psfrag{b}[c][c]{\huge $\mu$}
   \resizebox{0.9\columnwidth}{!}{
   \includegraphics{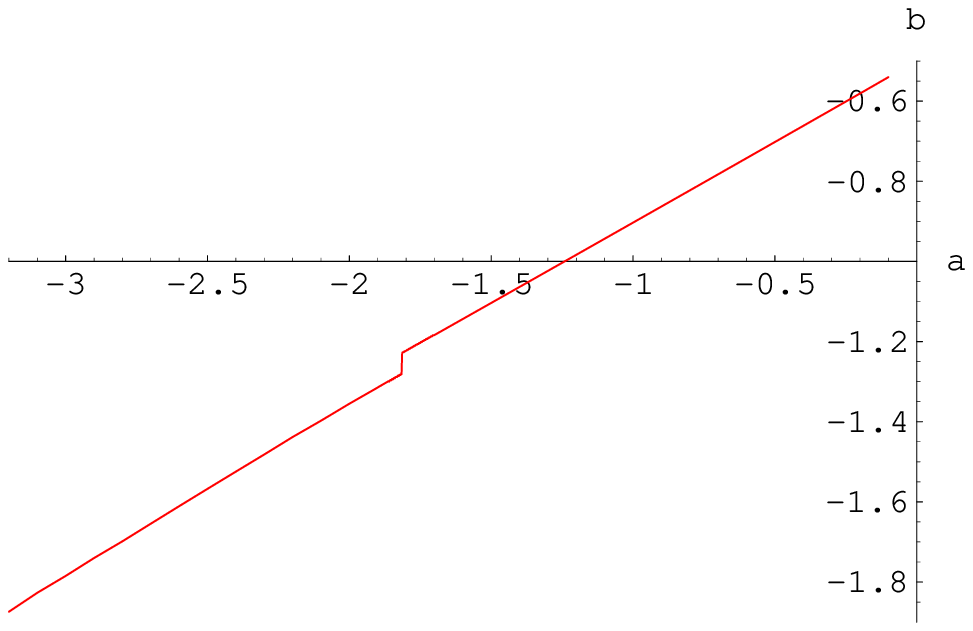}}
  \end{minipage}
  \begin{minipage}{58mm}
   \psfrag{a}[l][l]{\huge $U$}
   \psfrag{b}[c][c]{\huge $m$}
   \resizebox{0.9\columnwidth}{!}{
   \includegraphics{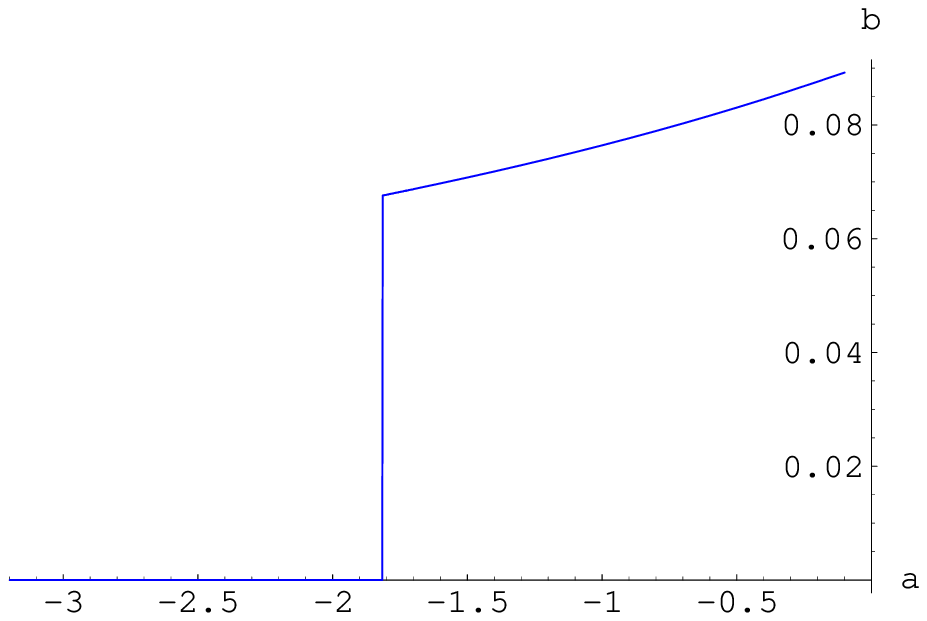}}
  \end{minipage}
  \begin{minipage}{58mm}
   \psfrag{a}[l][l]{\huge $U$}
   \psfrag{b}[c][c]{\huge $\Delta$}
   \resizebox{0.9\columnwidth}{!}{
   \includegraphics{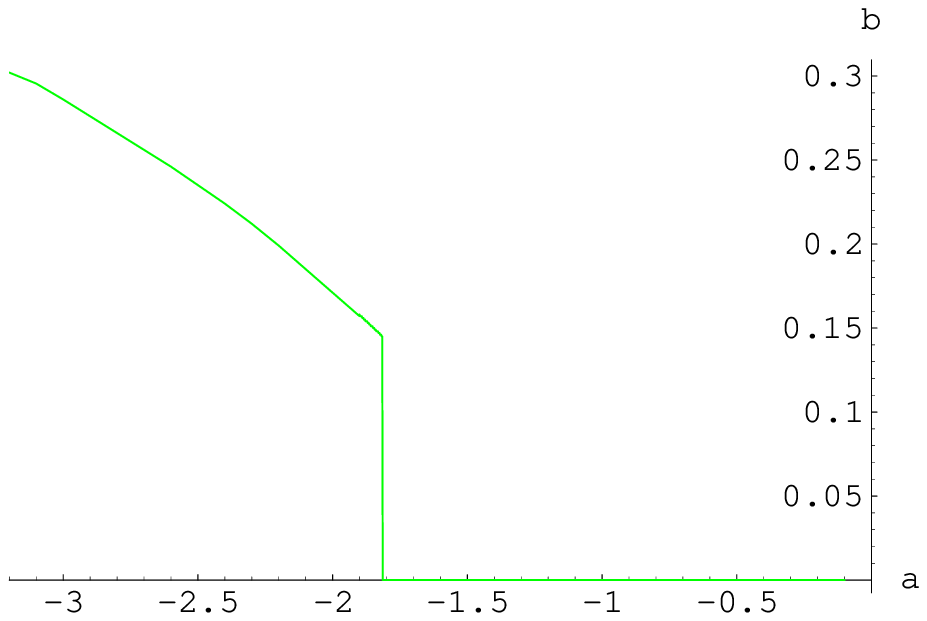}}
  \end{minipage}
 \end{center}
\caption{The parameters $\mu$, $m$ and $\Delta$ as a function of $U$ near half filling $n \approx 0.79$}\label{Gt7csg4Hio}
\end{figure*}

\begin{figure*}
 \begin{center}
  \begin{minipage}{58mm}
   \psfrag{a}[l][l]{\huge $U$}
   \psfrag{b}[c][c]{\huge $\mu$}
   \resizebox{0.9\columnwidth}{!}{
   \includegraphics{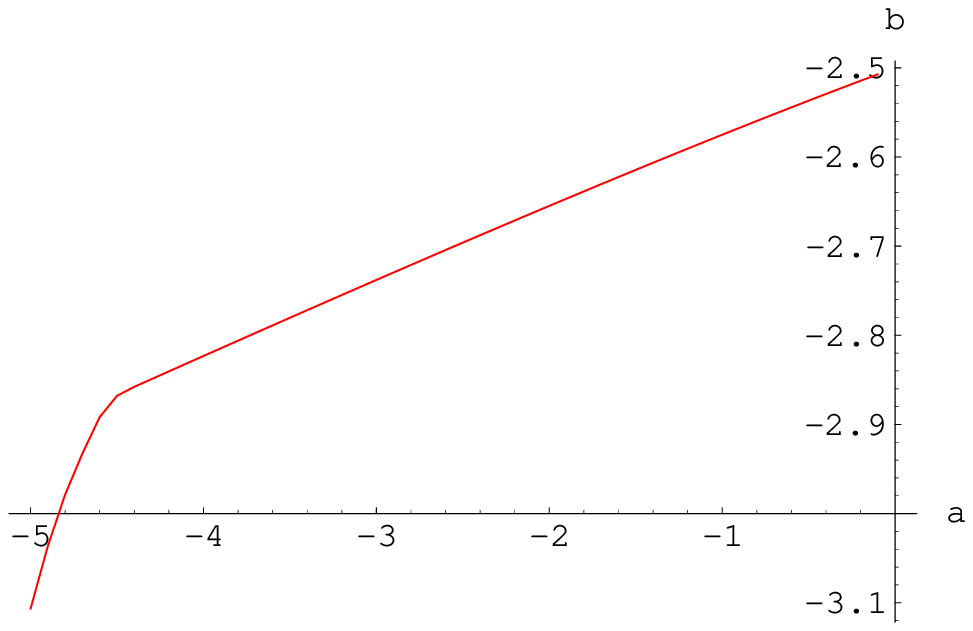}}
  \end{minipage}
  \begin{minipage}{58mm}
   \psfrag{a}[l][l]{\huge $U$}
   \psfrag{b}[c][c]{\huge $m$}
   \resizebox{0.9\columnwidth}{!}{
   \includegraphics{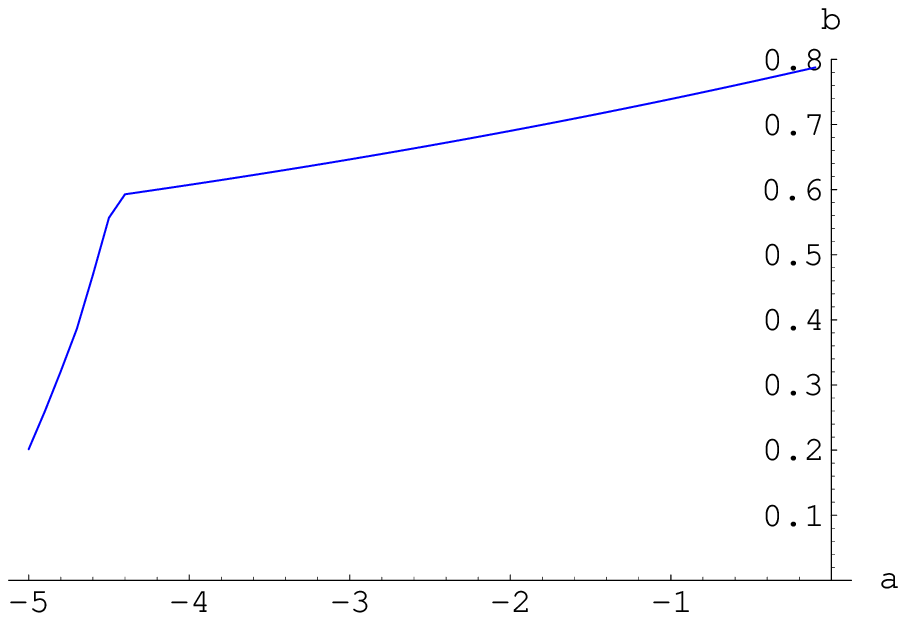}}
  \end{minipage}
  \begin{minipage}{58mm}
   \psfrag{a}[l][l]{\huge $U$}
   \psfrag{b}[c][c]{\huge $\Delta$}
   \resizebox{0.9\columnwidth}{!}{
   \includegraphics{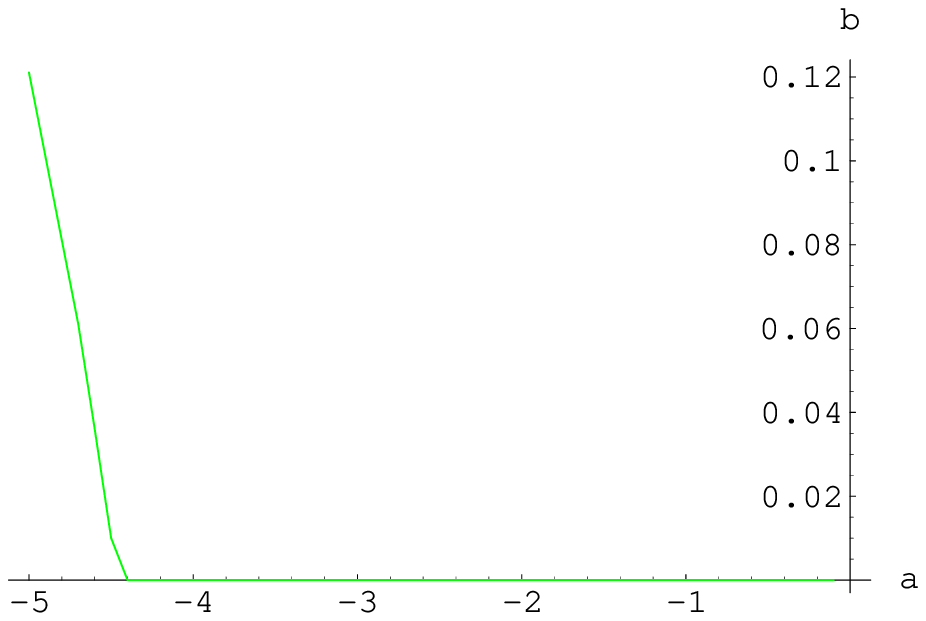}}
  \end{minipage}
 \end{center}
\caption{The parameters $\mu$, $m$ and $\Delta$ as a function of $U$ away half filling $n \approx 0.18$}\label{kLrdct4D2p}
\end{figure*}

In Figures \ref{Gt7csg4Hio} and \ref{kLrdct4D2p} we illustrate the behavior of the chemical potential $\mu$, the magnetization $m \equiv \frac{M}{n} = \frac{n_\uparrow - n_\downarrow}{n_\uparrow + n_\downarrow}$ and the absolute value of the superfluid order parameter $|\Delta|$ as a function of the interaction strength $U$ at different fixed occupation numbers $n=n_\uparrow + n_\downarrow$. We find that superfluid solutions exist only for attractive interaction strengths above a certain minimal value, $-U>U_{0}>0$. In contrast to systems with spin-independent hopping, we now also find thermodynamically stable magnetized superfluid solutions. Note that the trasitions occurring here are in general of second order.
\section{Summary and Discussion}\label{bd93scfguUa}
To summarize, we analyzed a generalized attractive-$U$ Hubbard model with (possibly spin-dependent) hopping amplitudes and a Zeeman-like magnetic field, which is relevant for ultracold quantum gases. In view of the spatial inhomogeneity due to the presence of a trap, we studied the model at Hartree-Fock level. For the special case of symmetric hopping we analytically demonstrated the uniqueness of the solution of Equations \eqref{bu7dfT53vk} and \eqref{bhYTD32sgT} at fixed order parameter $\Delta$. Hence we are certain to have detected all solutions of \eqref{bu7dfT53vk} - \eqref{o9d3acyU8fd}. We found that phase transitions at $B \neq 0$ are generally of first order. Within the local density approximation we showed that phase separation with an interesting shell structure occur if a parabolic potential is added to the Hamiltonian \eqref{v0skm59bb1}. For general spin-dependent hopping and $B=0$ we calculated the (spin-dependent) density of states. Interestingly, we found that, sufficiently far away from half filling, a superfluid magnetized phase occurs, which was absent in the phase diagram for spin-independent hopping.

Finally, we discuss possible extensions of our results. In this paper, we restricted consideration to
weak-coupling solutions at Hartree-Fock-LDA level. For our purposes, this restriction is meaningful, since
the Hartree-Fock-LDA approximation is expected to yield qualitatively correct results in the weak coupling
regime, provided spatial gradients are not too large; moreover, any other method would be virtually
impracticable in a spatially inhomogeneous system with a magneto-optical trap. Nevertheless, with an eye on
the future, it seems worthwhile to review possible alternatives.

First of all, it would probably be possible to drop the LDA-approximation and calculate the spatially
inhomogeneous Hartree-Fock solution for an optical lattice in a parabolical trap. However, even for a
two-dimensional lattice without spin-asymmetries ($B=0$ and $t_\uparrow = t_\downarrow$), this is known to be
computationally fairly expensive for realistic lattice sizes (with typically $10^3$ particles) \cite{freimuth:super}.

Secondly, it would of course be important to go beyond the Hartree-Fock approximation and include correlation effects. This could be done in principle by taking into account the dominant quantum fluctuations, which are described by second order Feynman diagrams in a perturbation expansion. In principle, such an expansion could be formulated also for spatially inhomogeneous systems with symmetry breaking, but the computational expense
of solving these equations self-consistently for a three-dimensional system of reasonable size would be tremendous. Here we just recall that it is known for translationally invariant systems \cite{dongen:futr93Ea,dongen:bgE58saB,dongen:gt8Yaqv4} in three or more dimensions, that the leading effect of quantum fluctuations is to renormalize Hartree-Fock results and reduce all energy scales (critical temperatures, gaps) by factors of typically three to four. For the calculations of the present paper this suggests that inclusion of second order diagrams would quantitatively but not qualitatively change the Hartree-Fock results.

Clearly it would, for the description of non-perturbative and strong-coupling effects, also be desirable to apply simulation methods (in particular Quantum Monte Carlo techniques) to the Hubbard models studied in this paper. However, a simulation of a three-dimensional system of physically interesting size at sufficiently low temperatures seems at present unfeasible. It also seems unlikely that approximations like Dynamical Mean Field Theory (DMFT) or its cluster extensions, which are known to be extremely helpful in translationally invariant systems in $d=2$ and $d=3$, are applicable to spatially inhomogeneous systems in a trap at {\em ultralow} temperatures, where superfluidity occurs. However, away from the symmetry-broken low-temperature phase, the DMFT has been demonstrated to be well applicable, e.g., to the description of spatially inhomogeneous Mott metal-insulator transitions \cite{helmes:bf5sc8H}.

For the time being, we conclude that we discovered several fascinating new phenomena in Hubbard-type models for asymmetric fermionic ultracold quantum gases. Specifically we found phase separation with an interesting shell structure for systems in a trap and also (in a translationally invariant model) a new superfluid magnetized phase. It would be important if the calculations presented here could be extended beyond the weak-coupling regime, using different methods, along the lines pointed out above. We hope and expect that our results can soon be compared to experiment.
\bibliographystyle{epj.bst}
\bibliography{masterbib.bib}
\end{document}